\documentclass[final,english]{bullsrsl}

\usepackage{natbib}

\usepackage[latin1]{inputenc}

\usepackage[T1]{fontenc}

\usepackage{graphicx}
\usepackage{float}
\usepackage{subfig}
\usepackage{hyperref}

\begin{document}

\title{Sector-based morphological analysis of four planetary nebulae \\(J 320, NGC 2610, A 6672 and PRTM 1) observed with UVIT/ASTROSAT}


  \author[affil={1}, corresponding]{Anisha}{Hazra}
  \author[affil={2}]{Ranjan}{Kumar}
  \author[affil={1}]{Sonika}{Piridi}
  \author[affil={1}]{Ananta}{C. Pradhan}
  

\affiliation[1]{Department of Physics \& Astronomy, National Institute of Technology, Rourkela, Odisha- 769008, India}
\affiliation[2]{Department of Physics, U. R. College, Rosera, Samastipur,  A constitutent unit of Lalit Narayan Mithila University, Darbhanga, Bihar$-$848210, India}


\correspondance{anisha.h1998@gmail.com}


\maketitle

\begin{abstract}
Four planetary nebulae (PNe), Jonckheere 320, NGC 2610, A 6672, and PRTM 1, have been imaged in the far-ultraviolet (FUV) filter F148W ($\lambda_{\mathrm {eff}} = 1481\, \mathring{A} $) using the Ultraviolet Imaging Telescope (UVIT) onboard ASTROSAT. The FUV images reveal faint extended emission structures surrounding the bright central star. A novel sector-based structural analysis technique was employed to examine the detailed FUV morphologies of PNe. In this method, sector strengths were calculated for each nebula and used to classify their structures. Our results show that Jonckheere 320 exhibits a polypolar morphology, while NGC 2610 and PRTM 1 display spherical structures. In contrast, A 6672 shows irregular or elliptical features in the FUV.
 
\end{abstract}

\section{Introduction}

Planetary nebulae (PNe) represent a short ($\sim 10^{4}-3\times10^{4}$ years) but critical phase in the late evolution of low- to intermediate-mass stars (1$-$8 M$_{\odot}$). During this stage, they eject their outer layer in a series of mass loss events, creating expanding shells of gas and dust. The exposed extremely hot core emits intense ultraviolet (UV) radiation, which ionizes the expelled circumstellar material, producing the luminous nebular shell \citep{Kaler1997}. PNe display a remarkable diversity of morphologies ranging from spherical and elliptical to bipolar and multi-polar, shaped by the complex interplay of stellar winds, binarity, magnetic fields, stellar rotation, and asymmetric mass loss \citep{BalickFrank2002,DeMarco2009,Zijlstra2015}. Their gaseous envelopes, composed of mainly hydrogen and helium but enriched with nucleosynthetic products such as carbon, nitrogen, and neon, and emit strong recombination and forbidden emission lines (e.g., H$\alpha$, H$\beta$, He,II, [O,III], and [N,II]), while molecular species such as H$_2$, CO, and polycyclic aromatic hydrocarbons (PAHs) contribute to infrared emission and play a key role in the chemical enrichment of the interstellar medium \citep{Aller1983,OsterbrockFerland2006}.
 
Multi-wavelength observations, particularly in the UV are essential because the hot central stars of PNe emit bulk of their energy at high temperatures, peaking in UV regime. This energetic radiation ionizes the surrounding ejecta and produces high-excitation emission lines such as N IV], C IV, O III], etc., and low-excitation emission lines such as O IV], N III, etc., that are weak or absent in the optical \citep{Pradhan2019}. Consequently, UV observations are critical for accurately constraining the central star properties, ionization structure, chemical abundances, and nebular kinematics, and for detecting faint halos and molecular fluorescence \citep{Kwitter2022}. The Galaxy Evolution Explorer (GALEX) has provided extensive UV imaging of a large number of PNe. These data have been compiled into comprehensive catalogs and used to characterize UV morphology of PNe \citep{Gomez2023, Pradhan2019, Bianchi2018}. The angular dimensions and overall morphology of PNe are crucial for investigating their various physical properties. 

The Ultraviolet Imaging Telescope (UVIT) onboard India's AstroSat satellite delivers high-resolution imaging in the far-UV (FUV: 130$-$180 nm) and near-UV (NUV: 200$-$300 nm) bands, with an angular resolution of better than 1.8 arcsec. This capability enables detailed studies of a wide range of astrophysical sources \citep{Piridi2024}, including PNe. UVIT reveals intricate morphological structures, UV-bright emission mechanisms and extended halos or faint outer envelopes that are often invisible or poorly resolved at other wavelengths. Previous observations of PNe such as NGC 40 and NGC 6302 have allowed detailed analyses of their hot central stars and their interaction with the surrounding interstellar medium \citep{Rao2018ngc40, KR2018}. In the present work, we perform a morphological analysis of four PNe (J 320, NGC 2610, A 6672, and PRTM 1) using FUV images obtained with UVIT.

\section{Data Analysis}

\begin{figure}[htb!]
    \hfill
\begin{center}
    \begin{minipage}[t]{0.24\textwidth}
        \includegraphics[width=\linewidth]{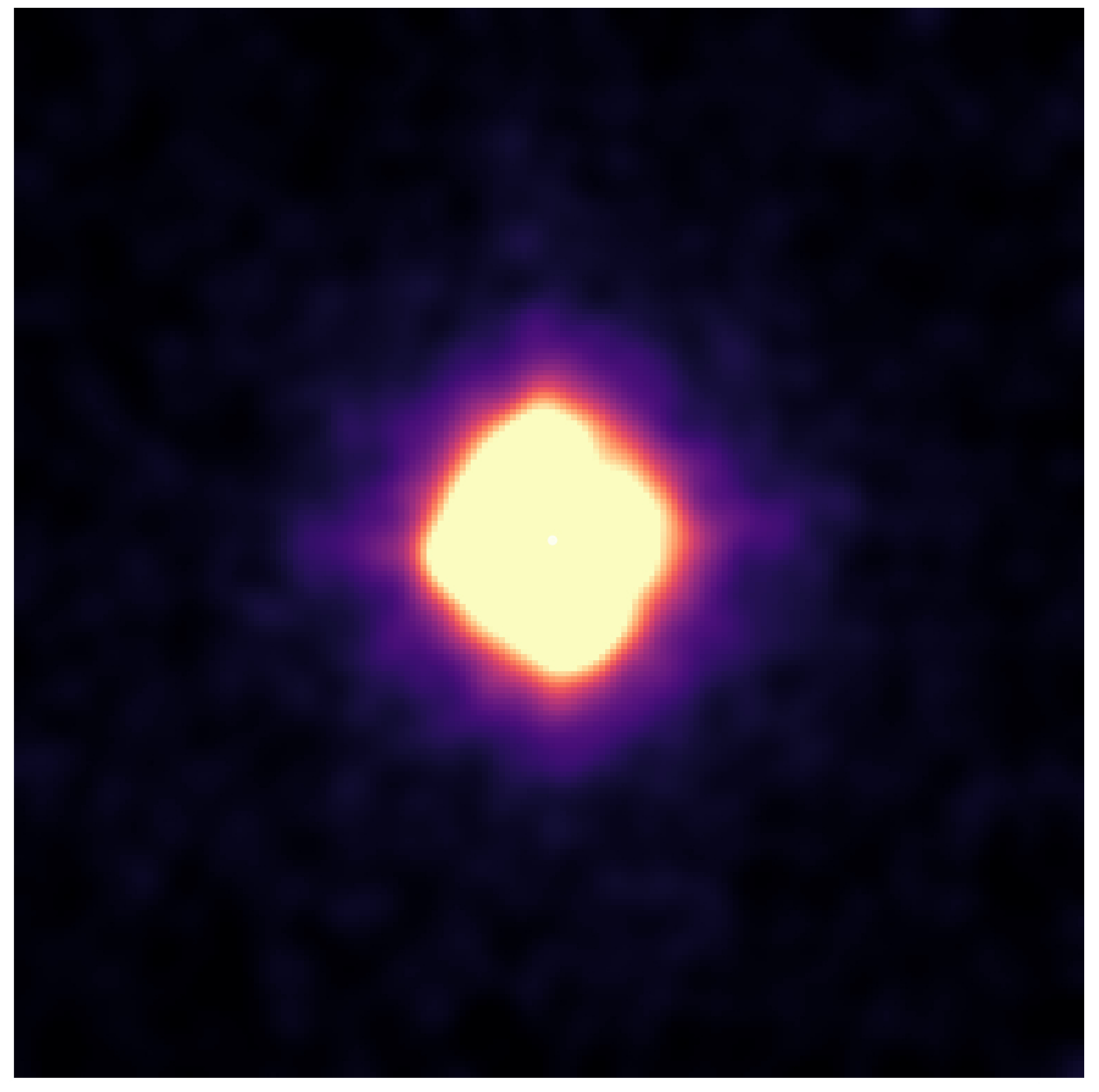}\\
        \centering J 320
    \end{minipage}
    \hfill
    \begin{minipage}[t]{0.24\textwidth}
        \includegraphics[width=\linewidth]{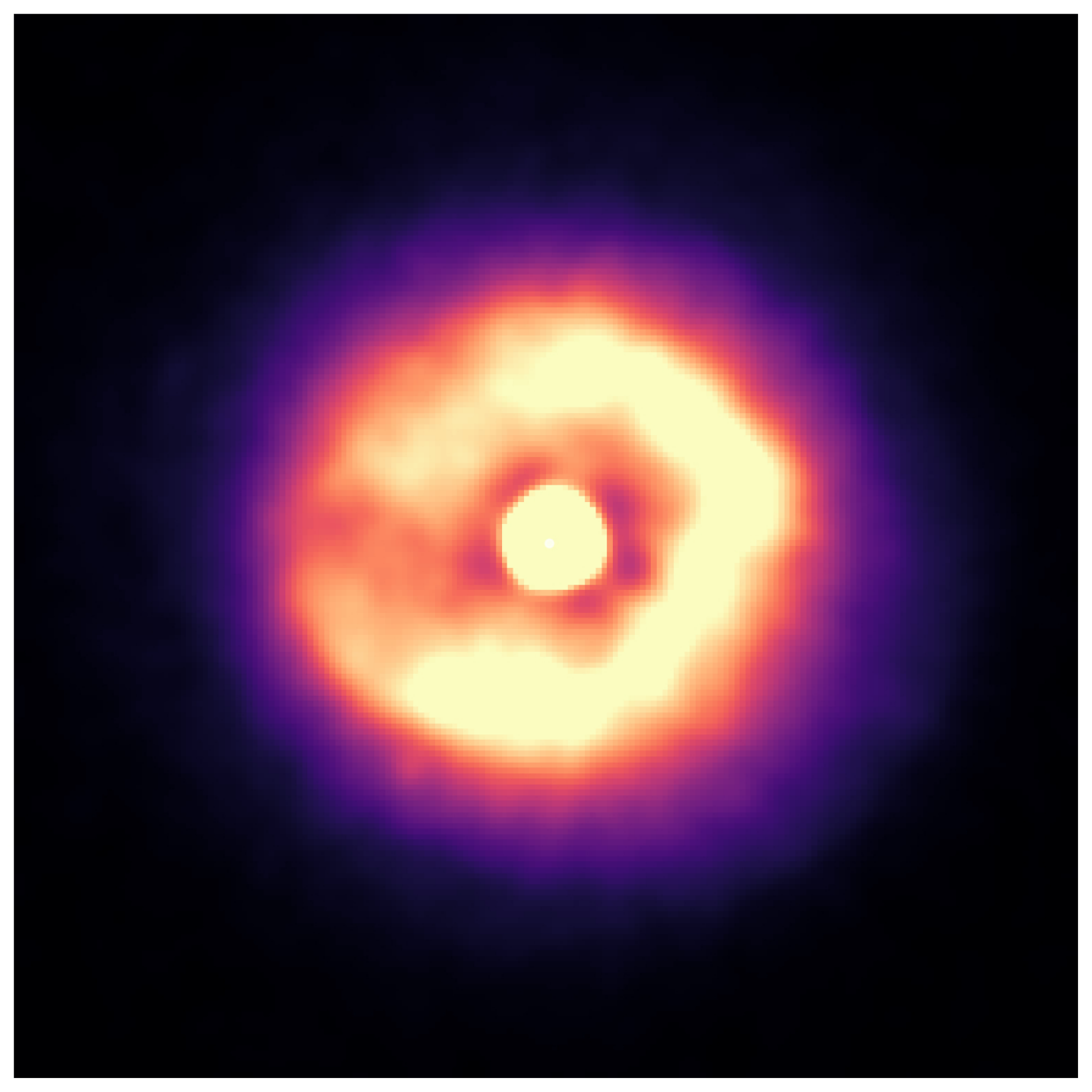}\\
        \centering NGC 2610
    \end{minipage}
    \hfill
    \begin{minipage}[t]{0.24\textwidth}
        \includegraphics[width=\linewidth]{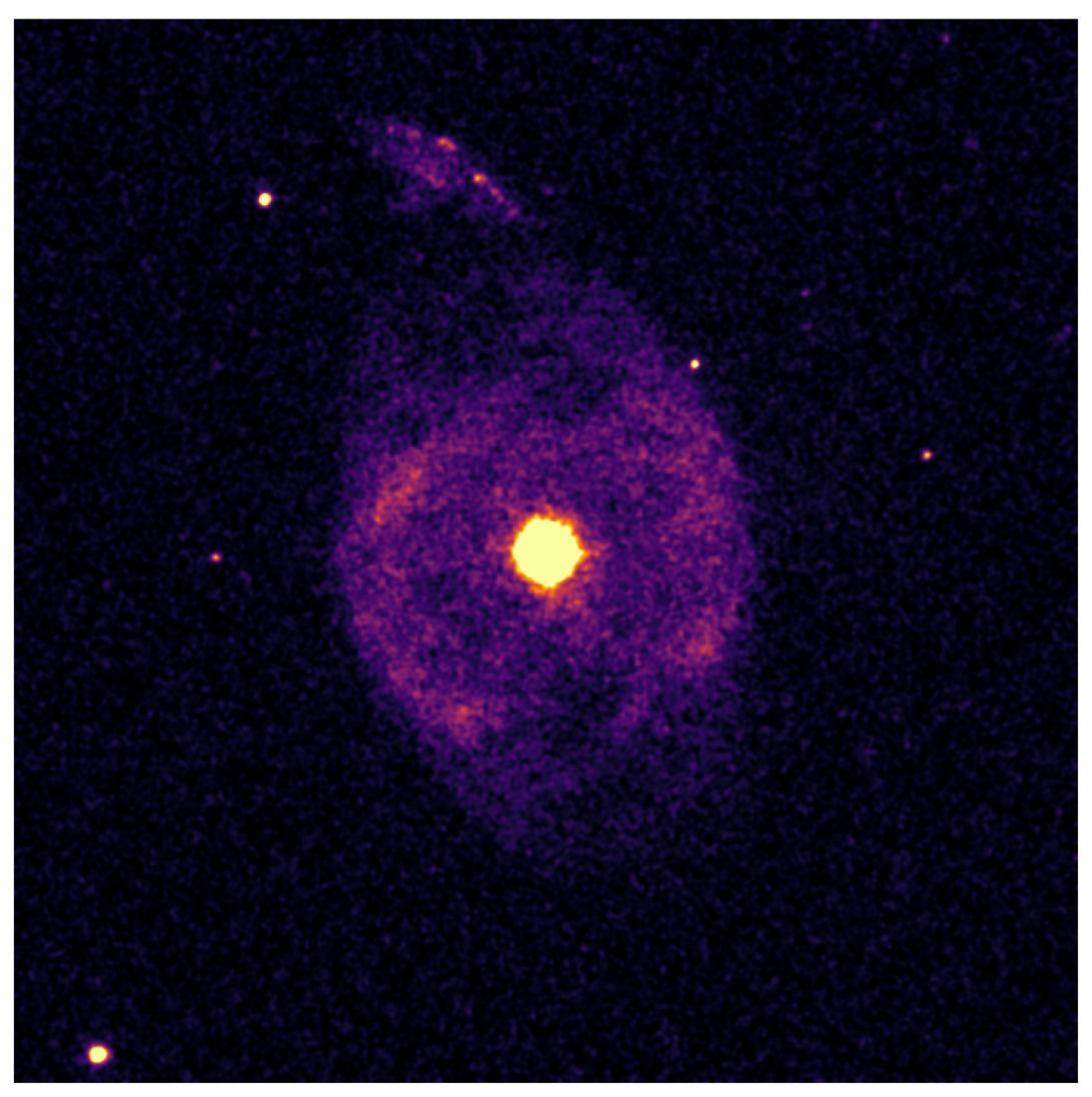}\\
        \centering A 6672
    \end{minipage}
    \hfill
    \begin{minipage}[t]{0.24\textwidth}
        \includegraphics[width=\linewidth]{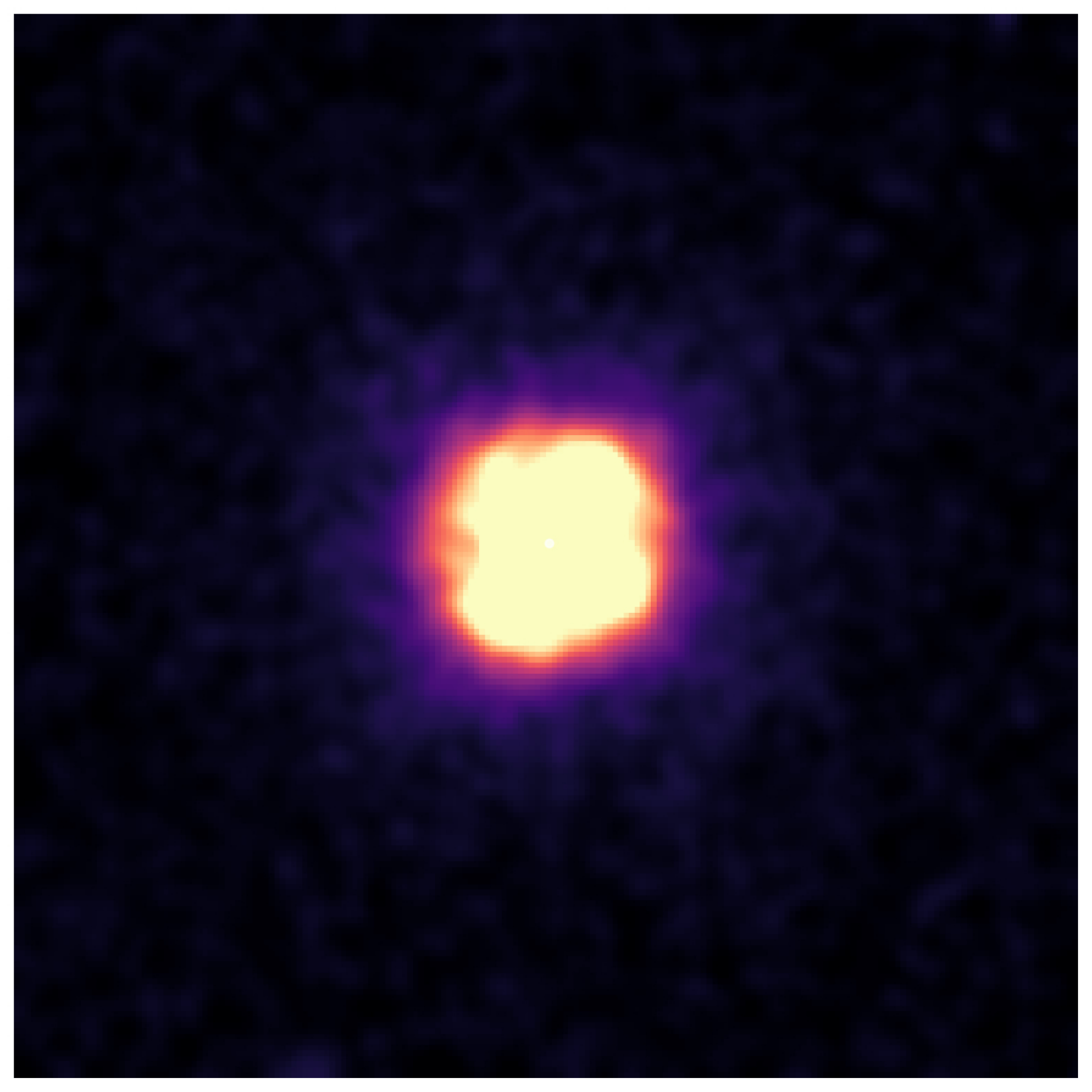}\\
        \centering PRTM 1
    \end{minipage}
     \caption{FUV ($\lambda_{\mathrm {eff}} = 1481\, \mathring{A} $) images of the PNe J 320, NGC 2610,  PRTM 1 and A 6672. The plasma colormap highlights the relative intensity distribution in false color.
      }
    \label{fig:4pne}
\end{center}
\end{figure}

All the four PNe were observed using the FUV CaF$_2$ (F148W) filter ($\lambda_{\mathrm {eff}} = 1481\, \mathring{A} $) of the UVIT/ASTROSAT telescope, spanning multiple orbits and imaging frames. The raw data were processed into science-ready images using the CCDLAB pipeline \citep{postma2017, Postma2021}. 
Point-spread function (PSF) photometry was performed using the \textit{PHOTUTILS} package, together with Astropy, in a Python-based data analysis framework \citep{Photutils2016}. The centroids of the central star of PNe (CSPNe) in the UVIT images were determined by fitting a Moffat-2D profile to the background-subtracted emission, with the fit refined in localized regions to minimize contamination. FUV images of the PNe are shown in Fig.~\ref{fig:4pne}. In these images, darker hues (black to deep purple) represent the lowest count rates and background emission, intermediate hues (purple to red/orange) indicate moderate surface brightness levels and the brightest regions of PNe, including the inner shell and the central emission peak, appear in yellow to white-yellow hues, corresponding to the highest observed flux. The photometric parameters such as exposure times, count rates, flux, and angular size of all PNe are provided in Table \ref{tab:pn_parameters}.

\begin{table}
\footnotesize
\centering
\begin{minipage}{160mm}  
\centering
\caption{UVIT observation details and photometric parameters of the four PNe}
\label{tab:pn_parameters}
\end{minipage}
\begin{tabular}{ccccccc}
\hline
\textbf{PN} &
\textbf{Obs. Date \& Time} &
\textbf{Exposure Time} &
\textbf{Count Rate} &
\textbf{Flux } & \textbf{Angular size} \\
 Name &  (UTC) &  (s) & (CPS)  & ($10^{-13}$ erg s$^{-1}$ cm$^{-2}$\,\AA$^{-1}$) & (arcsecs)  \\
  \hline
 J 320  & 25-08-2024  & 3526.393 & $30.867 \pm 0.030$ & $0.954 \pm 0.009$  & $25.2'' \times 22.52''$ \\
 & 11:05:13 & & & & \\
NGC 2610  & 14-11-2023 & 3459.655 & $117.068 \pm 0.018$ & $3.617 \pm 0.034$ & $35.0''\times 34.20''$\\
 & 20:42:45 & & & & \\
A 66\,72 & 15-04-2024 & 3823.773 & $41.270 \pm 0.008$ & $1.275 \pm 0.012$ & $120.0''\times 236.0''$ \\
 & 07:05:20 & & & & \\
PRTM 1   & 15-04-2024 & 3451.052 & $30.862 \pm 0.037$ & $0.954 \pm 0.009$ & $21.26'' \times 21.68''$ \\
 & 18:15:40 & & & & \\
\hline
\end{tabular}
\end{table}

\section{Methodology}

We introduce a novel sector-based pair strength modeling technique designed to quantitatively characterize the morphological properties of four PNe Jonckheere 320, NGC 2610, A 6672, and PRTM 1 observed in FUV with the UVIT. To ensure reliable detection of faint nebular emission, a careful global background subtraction was applied by averaging the signal from 10 source-free regions distributed across each image, such that the residual emission contains only stellar and nebular contributions. First, the stellar signal was modeled using a Moffat point-spread function (PSF), $I(r) = I_{0}\left[1 + \left(\frac{r}{\gamma}\right)^{2}\right]^{-\alpha}$, where $I_{0}$ is the background-subtracted central intensity and the shape parameters $\gamma$ and $\alpha$ were empirically determined from bright, isolated field stars in the same UVIT images. This PSF represents the expected radial decline of a pure point source in the absence of extended emission. Any observed flux exceeding the PSF model at a given radius was interpreted as a genuine nebular excess emission.

All the PNe were divided into $N=12$ equal angular sectors, each spanning $30^{\circ}$ in azimuth and centered on the nebula's centroid. For each angular sector, radial profiles were computed in concentric annuli of  five-pixel width. The observed flux, $F_{\rm obs}$ was compared directly with the PSF model to calculate the fractional excess flux, 
$f(r,\theta) = \frac{F_{\rm obs}(r,\theta) - F_{\rm PSF}(r)}{F_{\rm PSF}(r)}$, 
which isolates extended nebular emission beyond wings of the stellar PSF. 

These fractional excess profiles were radially integrated to obtain a single quantitative measure of extended emission per sector.

To identify large-scale axial symmetries in the nebular emission, the sectoral strengths were combined into opposite-sector pairs. For a total of $N=12$ sectors, each sector was paired with its diametrically opposite counterpart. For example, sector 1 ($0^\circ - 30^\circ$) was paired with sector 7 ($180^\circ - 210^\circ$), sector 2 ($30^\circ - 60^\circ$) with sector 8 ($210^\circ - 240^\circ$), and so on. The pair strength of each opposite-sector pair was defined as the sum of the values of the integrated strengths of the two opposing sectors. 
These pair strengths were then normalized by the total excess signal to obtain fractional pair contributions to represent the relative importance of each symmetry axis in shaping the nebula. This normalization allows a direct comparison between different nebulae and ensures that the sum of all pair fractions equals unity.

\section{Discussion and Conclusion}
\begin{figure}[htb!]
    \hfill
\begin{center}
    \begin{minipage}[t]{0.24\textwidth}
        \includegraphics[width=\linewidth]{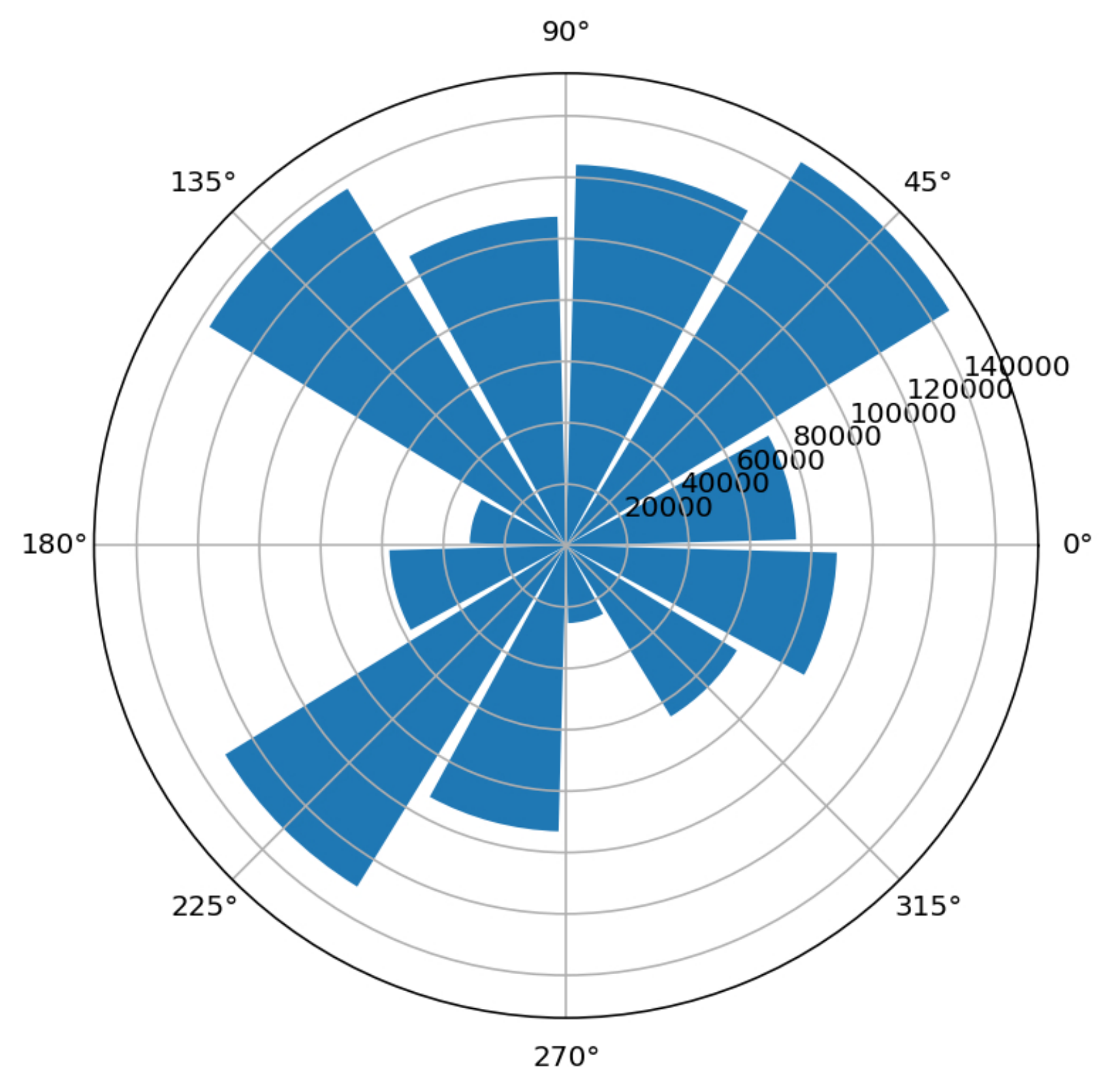}\\
        \centering J 320
    \end{minipage}
    \hfill
    \begin{minipage}[t]{0.24\textwidth}
        \includegraphics[width=\linewidth]{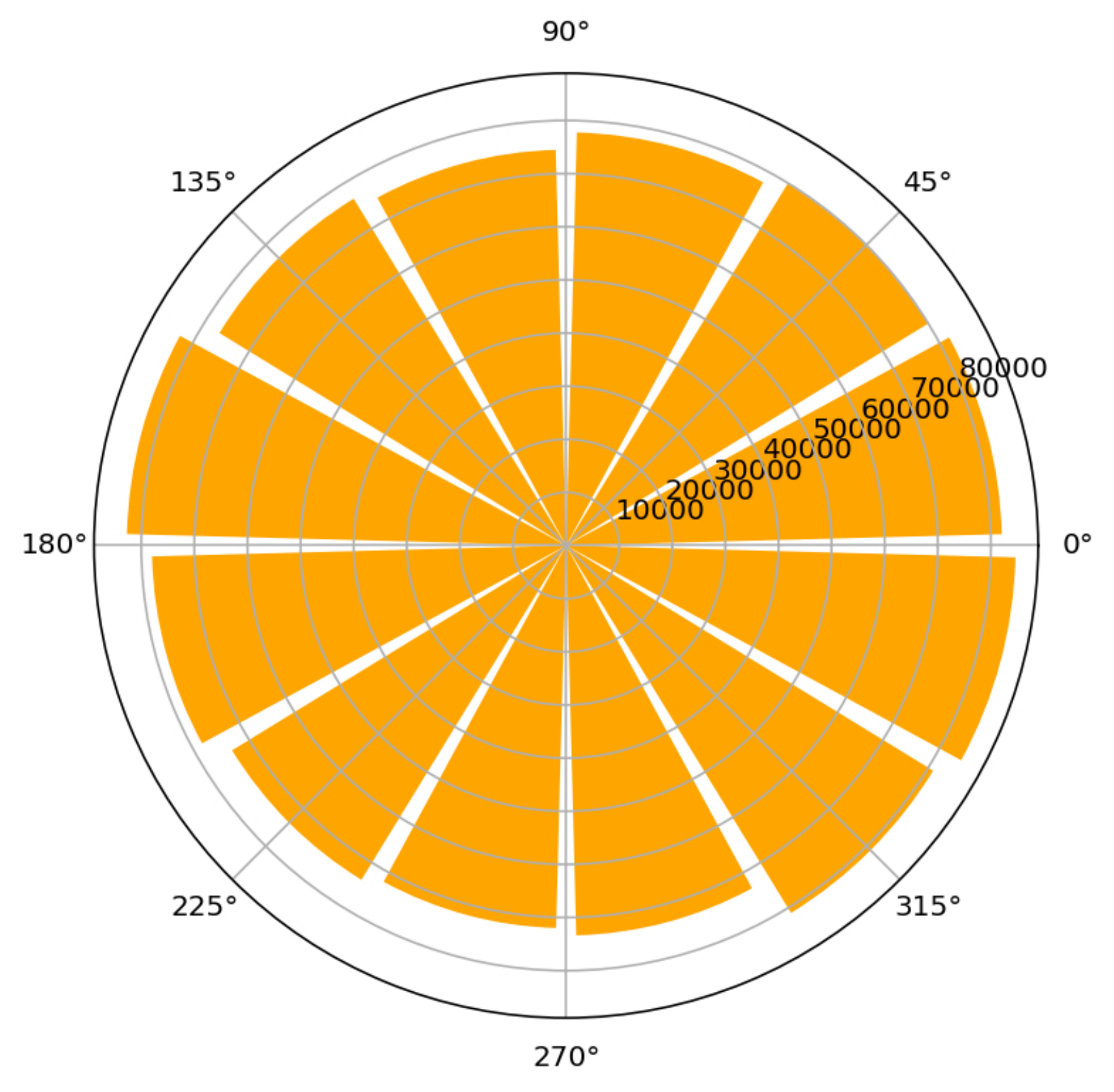}\\
        \centering NGC 2610
    \end{minipage}
    \hfill
    \begin{minipage}[t]{0.24\textwidth}
        \includegraphics[width=\linewidth]{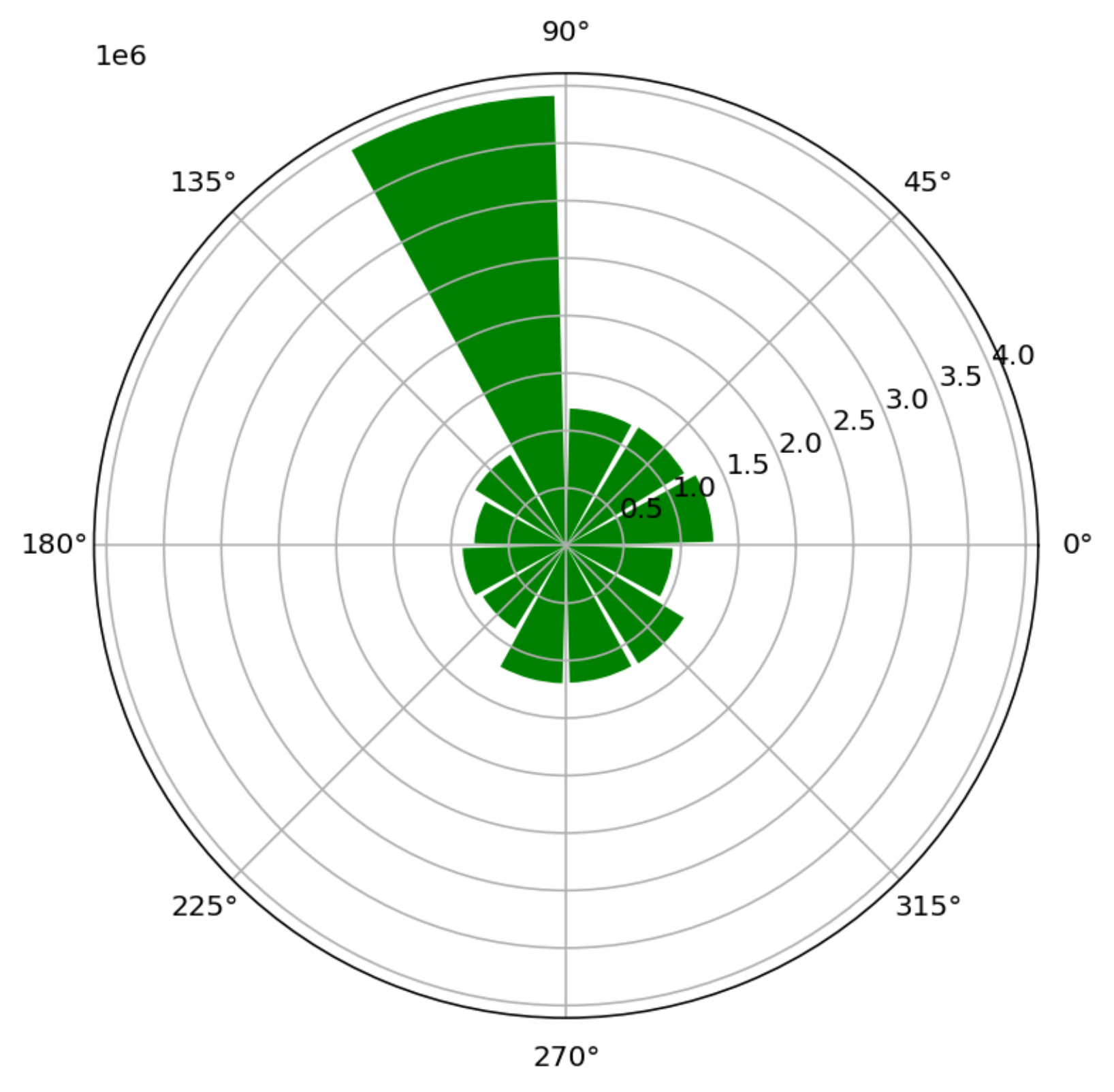}\\
        \centering A 6672
    \end{minipage}
    \hfill
    \begin{minipage}[t]{0.24\textwidth}
        \includegraphics[width=\linewidth]{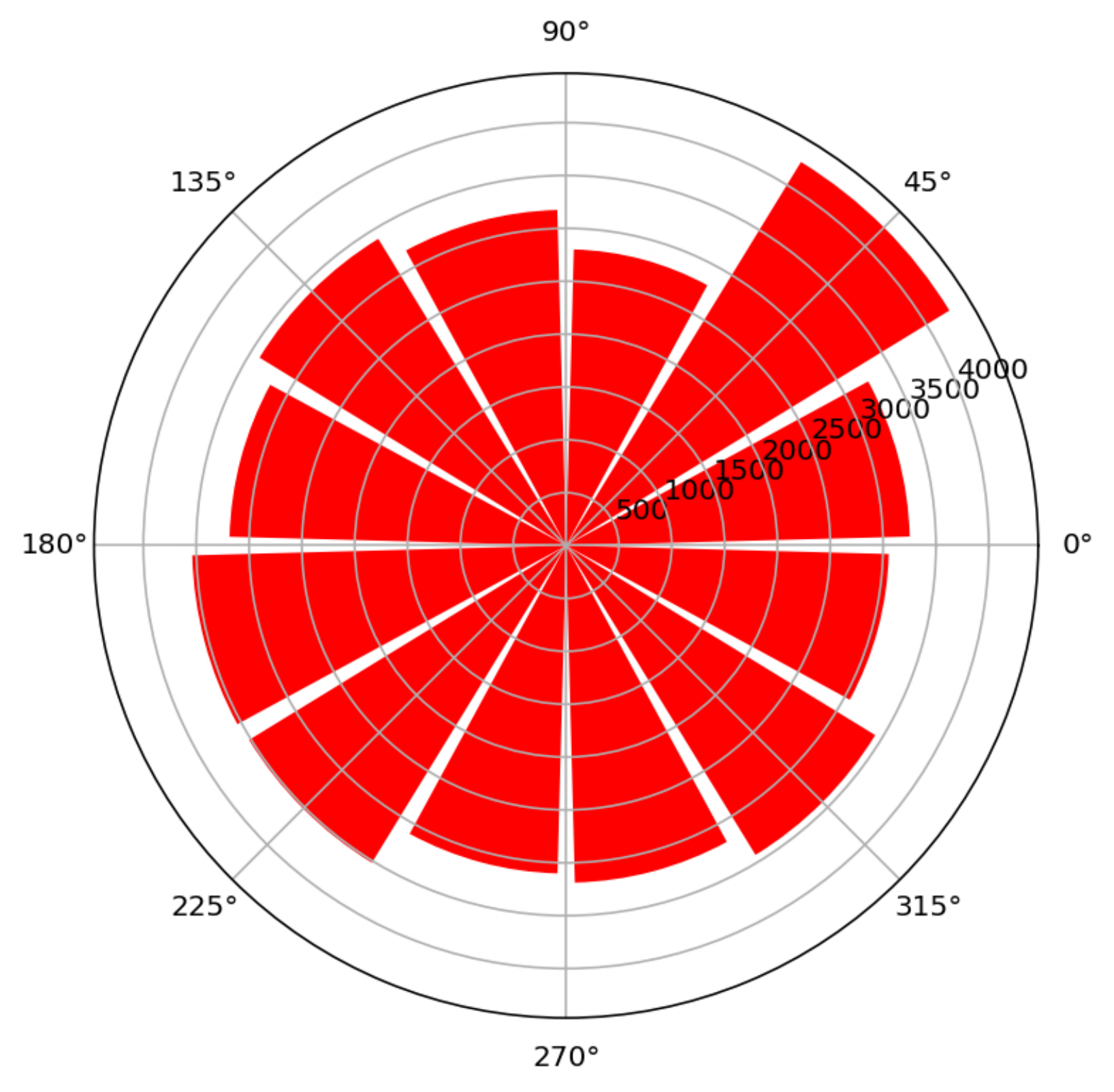}\\
        \centering PRTM 1
    \end{minipage}
    \caption{Polar bar plots show the integrated fractional excess emission strength in 12 angular sectors for all PNe, highlighting the dominant directions of extended nebular emission.}
    \label{fig:Sector Strength}
\end{center}
\end{figure}

The fractional strengths of opposite sector pairs, along with the resulting morphological classifications based on this method, are presented in Table \ref{tab:pair_strengths}, while polar bar plots of the sector strengths are shown in Fig. ~\ref{fig:Sector Strength}. For the first time, we provide a quantitative basis for the morphological classification of PNe using the fractional pair strength values listed in Table \ref{tab:pair_strengths}. Specifically, a nebula is classified as bipolar if a single opposite-sector pair contributes $\geq 50\%$ of the total excess flux, indicating a dominant symmetry axis. It is classified as polypolar if two or more opposite-sector pairs each contribute between 20\% and 30\%, consistent with multiple lobe axes or episodic outflows. A spherical morphology is assigned when almost all pairs contribute nearly equally ($\sim 15-20\%$), with the standard deviation of all pair fractions being less than 0.1, reflecting near-isotropic emission. Finally, sources are classified as irregular/elliptical when no clear dominance or uniformity is observed, indicating complex or asymmetric structures.


\begin{table}
\centering
\begin{minipage}{160mm}  
\centering
\caption{Fractional opposite-sector pair strengths derived from sector-based excess-flux analysis for the four PNe. Each pair corresponds to diametrically opposite angular sectors separated by $180^{\circ}$.}
\label{tab:pair_strengths}
\end{minipage}
\begin{tabular}{cccccccc}
\hline
\textbf{PN} & \textbf{Pair 1} & \textbf{Pair 2} & \textbf{Pair 3} & \textbf{Pair 4} & \textbf{Pair 5} & \textbf{Pair 6} & \textbf{Morphology}  \\
\hline
Jonckheere 320 & 0.123 & 0.256 & 0.202 & 0.123 & 0.187 & 0.120 & Polypolar \\
NGC 2610       & 0.171 & 0.164 & 0.160 & 0.158 & 0.168 & 0.179 & Spherical \\
A 6672          & 0.140 & 0.132 & 0.153 & 0.328 & 0.137 & 0.111 & Elliptical \\
PRTM 1         & 0.170 & 0.195 & 0.148 & 0.159 & 0.171 & 0.156 & Spherical \\
\hline
\end{tabular}
\end{table}

Based on the above classification scheme, we discuss four PNe Jonckheere 320, NGC 2610, A 6672, and PRTM 1 as follows:
\begin{itemize}
\item \textbf{J 320 (PN\,G190.3$-$17.7)}: For this planetary nebula, the nebular emission is distributed across multiple axes rather than being dominated by a single bipolar pair. Two pairs of opposite-sectors contribute substantially to the total flux, with fractional contributions of $\sim$ 0.256 and 0.202, respectively. This morphology is a hallmark of polypolar PNe. In addition, the presence of multiple pairs of point-symmetric, high-velocity knots surrounding the main lobes is consistent with relics of episodic, precessing jets launched at different epochs and orientations, as proposed for multipolar nebula \citep{Harman2004}.

\item \textbf{NGC\,2610 (PN\,G239.6+13.9)}: The opposite-sector pair fractions for NGC 2610 (Table \ref{tab:pair_strengths}) exhibit very small dispersion, indicating an almost perfectly isotropic distribution of excess flux and the absence of any preferred symmetry axis. This high degree of uniformity is characteristic of the rare round PNe and matches expectations for slowly evolving AGB-star ejecta that have not been strongly shaped by binary interactions or magnetic fields. High-resolution imaging and kinematic studies in the literature likewise describe NGC 2610 as a smooth, symmetric, high-excitation shell lacking low-ionization knots or filaments \citep{Harrington2006}, in excellent agreement with our results.

\item \textbf{A 6672 (PN\,G059.7$-$18.7)}: This is an evolved elliptical planetary nebula with exhibiting a weakly defined symmetry axis, rather than a strongly bipolar morphology. The observed structure may be influenced by projection effects  \citep{Schwarz1992,Corradi1995}. As shown in Fig.~\ref{fig:Sector Strength}, there is a clear but non-dominant preferred axis, with one pair of opposite sectors contributing a significantly larger fraction of the excess emission ($\sim$0.32) compared to the other pairs ($\sim 0.11 - 0.15$) (Table \ref{tab:pair_strengths}). This indicates anisotropic expansion in the absence of a single dominant bipolar axis. Overall, the emission pattern points to mild elongation combined with irregular clumpiness, rather than a canonical bipolar morphology.

\item \textbf{PRTM\,1 (PN\,G243$-$37.1)}: For PRTM 1, our sector-based analysis (Fig.~\ref{fig:Sector Strength}) shows nearly uniform pair fractions ($\sim$0.15$-$0.19) with no dominant axis (Table \ref{tab:pair_strengths}). This reinforces the overall spherical symmetry of the system. Morphology, the surrounding planetary nebula is almost perfectly spherical and has several shells. According to \citep{Boffin2012}, it lies in a high-ionization environment characterized by thin inner arcs along the major axis, while lacking low-ionization structures such as jets, knots, or fliers. The absence of these features indicate minimal asymmetrical mass ejection. Taken together, these results indicate that PRTM~1 is best described as a nearly spherical planetary nebula with minor asymmetries outflows. 
\end{itemize}

 Overall, we have studied the UV morphology and sizes of four PNe observed with UVIT. We employed a novel sector-based pair-strength modeling technique to characterize their structures. Our results show that Jonckheere 320 exhibits a polypolar morphology, NGC 2610 and PRTM 1 are spherical, and A 6672 displays an elliptical morphology.
In the future, we aim to interpret the physical drivers behind these diverse nebular morphologies through lobe-specific mass-loss profile modeling, supported by detailed kinematic analyses and multi-wavelength observations.

\begin{acknowledgments}
 We acknowledge the use of data from the AstroSat mission of the Indian Space Research Organisation (ISRO), archived at the Indian Space Science Data Centre (ISSDC). ACP and SP acknowledge the support of Indian Space Research Organisation (ISRO) under AstroSat archival Data utilization program (No.DS 2B-13013(2)/1/2022-Sec.2).

\end{acknowledgments}

\begin{furtherinformation}

 \orcid{0009-0009-8899-169X}{Anisha}{Hazra}
 \orcid{0000-0002-9273-6001}{Ranjan}
 {Kumar}
 \orcid{0000-0003-3083-181X}{Sonika}{Piridi}
 \orcid{0000-0001-5808-0654}{Ananta}{C. Pradhan}

\begin{authorcontributions}
All authors have contributed equally.
\end{authorcontributions}

\begin{conflictsofinterest}
The author declares that there is no conflict of interest.
\end{conflictsofinterest}

\end{furtherinformation}



%

\bibliographystyle{bullsrsl-en}

\bibliography{extra}


\end{document}